\documentclass{article}
\usepackage{epsfig}

\date{\today}

\begin{document}
\begin{center}

{\Large Rotating Yang-Mills dyons in anti-de Sitter spacetime}
\vspace{0.6cm}
\\
Eugen Radu
\\
{\small
\it Albert-Ludwigs-Universit\"at, Fakult\"at f\"ur Physik, Hermann-Herder-Stra\ss e 3
\\
 Freiburg D-79104, Germany}
\end{center}
\begin{abstract}
We construct new axially symmetric solutions of 
SU(2) Yang-Mills theory in a four dimensional anti-de Sitter spacetime.
Possessing nonvanishing nonabelian charges, these regular configurations have also a
nonzero angular momentum.
Numerical arguments are also presented for the existence of 
rotating solutions of  
Einstein-Yang-Mills equations in an asymptotically anti-de Sitter spacetime.
\end{abstract}

\noindent{\textbf{Introduction.--~}}
Given its important role in string theory, 
anti-de Sitter (AdS) spacetime has recently attracted huge interest. 
As proven by some authors \cite{Bjoraker:2000qd, Winstanley:1998sn}, 
even the simple spherically symmetric $SU(2)$ Einstein-Yang-Mills (EYM) 
system with a negative cosmological constant $\Lambda$
presents some surprising results.
A variety of well known features of asymptotically flat self-gravitating nonabelian
solutions are not shared by their AdS counterparts. 
First, there is a continuum of regular and black hole 
solutions in terms of the adjustable shooting parameters
that specifies the initial conditions at the origin or at the event horizon,  
rather then discrete points.
The spectrum has a finite number of continuous branches.
Secondly, there are nontrivial solutions stable against spherically symmetric 
linear perturbations, corresponding 
to stable monopole and dyon configurations.
The solutions are classified 
by non-Abelian electric and magnetic charges 
and the ADM mass.
When the parameter $\Lambda$ approaches zero, 
an already-existing branch of monopoles and dyon solutions 
collapses to a single point in the moduli space \cite{Hosotani:2001iz}.
At the same time new branches of solutions emerge.

An interesting physical question is whether the known static nonabelian solutions can be
generalized to include an angular momentum.
In a EYM theory without a cosmological constant,
slowly rotating charged solutions are known to
exist \cite{Brodbeck:1997ek}. 
For regular configurations, 
the angular momentum and the electric charge are related.
The existence of these perturbative rotating solutions came as 
a surprise, given the experience with other solitonic solutions
\cite{Heusler:1998ec}. 
However, it is not clear
whether they can be extended to exact solutions \cite{Sudarsky:ty, VanderBij:2001nm}. 
Thus it may be important to consider
other types of asymptotics for a better understanding of this question.

Here we consider this problem in a AdS geometry for a $SU(2)$ field.
Although an analytic or approximate solution still appears to be intractable,
we present numerical arguments that a negative cosmological constant
could support finite energy nonabelian configurations 
with a nonvanishing angular momentum.
These regular solutions are, in nature, nontopological solitons and 
carry both angular momentum and electric charge.
Moreover, they are found to survive in the presence of gravity.

\noindent{\textbf{Axially symmetric ansatz and general relations.--~}}
We consider the Yang-Mills (YM) equations
\begin{equation}
\label{YM-eqs}
\nabla_{\mu}F^{\mu\nu}+ie[A_{\mu},F^{\mu\nu}]=0,
\end{equation}
in a fixed AdS background
\begin{eqnarray}
\label{AdS-metric}
ds^2=  \frac{d r^2}{1-\frac{\Lambda}{3}r^2}+
r^2 (d \theta^2+\sin ^2 \theta d\phi)^2 - (1-\frac{\Lambda}{3}r^2) dt^2,
\end{eqnarray}
where $e$ is the gauge coupling constant,
$F_{\mu \nu} = 
\partial_\mu A_\nu -\partial_\nu A_\mu + i e \left[A_\mu , A_\nu \right] $ and 
$A_{\mu} = \frac{1}{2} \tau^a A_\mu^a$.
We are interested in stationary, axially symmetric 
solutions of the $SU(2)$ YM configurations with a nonvanishing total angular momentum. 

For the time translation symmetry, we choose a natural gauge 
such that $\partial A/\partial t$=0.
However, a rotation around the $z-$axis 
can be compensated by a gauge rotation 
${\mathcal{L}}_\varphi A=D\Psi$ \cite{Forgacs:1980zs}, 
with $\Psi$ being a Lie-algebra valued gauge function.
Therefore we find
$ F_{\mu \varphi} = D_{\mu}W, $
where $W=A_{\varphi}-\Psi$.

The energy density of the solutions is given by the
$tt$-component of the energy momentum tensor $T_{\mu}^{\nu}$; integration
over all space yields their energy
\begin{eqnarray}\label{mass}
E =  \int 2Tr \{ -F_{\mu t} F^{\mu t}+\frac{1}{4} F_{\mu \nu} F^{\mu \nu} \}
 \sqrt{-g}d^3x .
\end{eqnarray}  
The electric part in the above relation can be expressed 
as a surface integral at infinity
\begin{eqnarray} 
\label{electric-mass1}
-E_e =\int Tr\{ F_{\mu t} F^{\mu t} \}\sqrt{-g}d^3x=
 \oint_{\infty} Tr \{A_t F^{\mu t} \} dS_{\mu}.
\end{eqnarray}
Thus, similar to the asymptotically flat case \cite{Sudarsky:ty},
a vanishing magnitude of the electric potentials at infinity $|A_t|$ 
implies a purely magnetic solution.
For the gauge invariant nonabelian charges we use the expressions \cite{Sudarsky:ty,
Corichi:2000dm, Kleihaus:2000kg}
\begin{eqnarray}
\label{charges}
Q_e=\frac{1}{4\pi}\oint_{\infty}|^{\ast}F|,~~~Q_M=\frac{1}{4\pi}\oint_{\infty}|F|,
\end{eqnarray}
where the vertical bars denote the Lie-algebra norm.
The total angular momentum of a solution is given by
\begin{eqnarray}
\label{J}
J&=&\int T_{\varphi}^{t}\sqrt{-g} d^{3}x
= \int 2Tr\{F_{r \varphi} F^{r t}
+F_{\theta \varphi} F^{\theta t}\} \sqrt{-g} d^{3}x,
\end{eqnarray}
and can be expressed as a surface integral at infinity \cite{VanderBij:2001nm}
\begin{eqnarray}
\label{totalJ}
J &=&\oint_{\infty}2Tr\{WF^{\mu t} \} dS_{\mu}.
\end{eqnarray}
The usual ansatz used 
when discussing axially symmetric YM configurations 
in spherical coordinates is
\begin{equation}
\label{gauge-ansatz}
A_\mu dx^\mu =
\frac{1}{2e} 
\left[ \tau^n_\phi 
\Big( \frac{H_1}{r} dr + \left(1-H_2\right) d\theta \Big)
-n \Big( \tau^n_r H_3 + \tau^n_\theta \left(1-H_4\right) \Big)
 \sin \theta d\phi 
+\left( \tau^n_r H_5 + \tau^n_\theta H_6 \right) dt
\right],  
\end{equation}
with the Pauli matrices $\vec \tau = ( \tau_x, \tau_y, \tau_z) $ and
$\tau^n_r = \vec \tau \cdot 
(\sin \theta \cos n \phi, \sin \theta \sin n \phi, \cos \theta)$,
$\tau^n_\theta = \vec \tau \cdot 
(\cos \theta \cos n \phi, \cos \theta \sin n \phi, -\sin \theta)$,
$\tau^n_\phi = \vec \tau \cdot (-\sin n \phi, \cos n \phi,0)$ \cite{foot-1}.
The six gauge field functions $H_i$  depend only on $r$ and $\theta$.
To fix the residual abelian gauge invariance we choose the gauge condition 
$ r \partial_r H_1 - \partial_\theta H_2 = 0$ \cite{Kleihaus:1997mn,Kleihaus:2000kg}.
The integer $n$ represents the winding number of the solutions.
For $n=1$ and $H_1=H_3=H_5=H_6=0$, $H_2=H_4=w(r)$ 
the spherically symmetric magnetic ansatz of ref.~\cite{Hosotani:2001iz} 
is recovered.

For the magnetic potentials we impose the familiar boundary conditions
$ H_{2}=H_{4}= 1,~H_{1}=H_{3}=0 $
 at the origin and 
$ H_2=H_4= w_0,~H_1=H_3=0 $
at infinity, where $w_0$ has an arbitrary value.
Given the parity reflection symmetry, we need to consider solutions 
only in the region $0 \leq \theta \leq \pi/2$.
The boundary conditions satisfied by 
the magnetic potentials along the axes ($\theta=0,\pi/2$) are
$
H_1=H_3=0,~\partial_\theta H_2= \partial_\theta H_4=0. 
$

For $\Lambda<0$, there are no boundary conditions to exclude a nonabelian solution
with nonzero electric potential.
In \cite{Radu:2001ij} numerical arguments have been presented for the existence of
axially symmetric YM solutions with nonvanishing magnetic and electric charges
and winding number $n>1$. These solutions generalize the $n=1$ spherically symmetric
dyons discussed in \cite{Bjoraker:2000qd}.
For that type of configurations, the electric potentials $H_5,H_6$
satisfy a set of boundary conditions 
inspired by the flat-space Yang-Mills-Higgs (YMH) dyons \cite{Hartmann:2000ja}.
For example, we imposed at infinity $H_{5}=V, H_{6}=0$, while at the origin
$H_5=H_6=0$. 
Nevertheless,
we find that these dyon configurations
posses a vanishing total angular momentum $J$ \cite{foot0}.

However, already for a vanishing cosmological constant,
the above boundary conditions 
does not exhaust all possibilities even for $n=1$.
The monopole-antimonopole solutions (MAP) in a YMH theory \cite{Kleihaus:1999sx, Kleihaus:2000hx}
or the asymptotically flat
rotating EYM black holes discussed in \cite{Kleihaus:2000kg} are examples
of nonspherically symmetric configurations with unit winding numbers.
It is natural to expect that similar configurations will exist also in a AdS geometry. 
In what follows, we use for the electric potentials a set of boundary conditions 
inspired by the MAP configurations. 
We impose $H_5 \sin\theta + H_6 \cos\theta =0,~H_{5,r} \cos\theta - H_{6,r} \sin\theta = 0$
at the origin and 
$H_{5}=V \cos \theta,~H_{6}=V \sin \theta$
at infinity. 
The boundary conditions on the symmetry axis ($\theta=0$) are 
$\partial_{\theta}H_5=H_6=0$, while for $\theta=\pi/2$ we impose 
$H_5=\partial_{\theta}H_6=0$.
In this letter we consider only solutions with the lowest winding number $n=1$.

The boundary conditions for the gauge field functions at infinity 
imply that the solutions have a magnetic charge $Q_M=|1-\omega_0^2|$.
The value of the electric charge can also be obtained from the asymptotics of the
electric potentials, since as $r \to \infty$, 
$H_5\sim\cos \theta \big(V+(c_1 \sin^2 \theta +c_2)/r\big),~
H_6\sim\sin \theta \big(V+(c_3 \sin^2 \theta +c_4)/r\big)$.

\noindent{\textbf{Solutions in a fixed AdS background.--~}}
Because the asymptotic structure of geometry is different, 
in an AdS spacetime one does not have to couple the YM system to scalar 
fields or gravity in order to obtain finite energy solutions.
Here the cosmological constant breaks the scale 
invariance of pure YM theory to give finite energy solutions \cite{Hosotani:2001iz,Radu:2001ij}.
For spherical symmetry, finite energy monopole configurations are found
for only
one interval of the parameter $b$ that specifies the initial conditions at the origin. 
Depending on the value of $b$, the gauge function $\omega(r)$ is nodeless or presents one node only.

We start by presenting rotating dyons solutions of YM equations in a 
four-dimensional AdS spacetime, gravity being regarded as a fixed
field.
Although being extremely simple, nevertheless this model appears to contain all 
the essential features of the gravitating solutions.
Also, it is much easier to solve the field equations in this case.

Subject to the above boundary conditions,
we solve the YM equations numerically.
The numerical calculations are performed by using the program
FIDISOL \cite{FIDISOL}, based on the iterative Newton-Raphson method.
To map spatial infinity to a finite value,
we employ the radial coordinate $\bar r = r/(r+c)$,
 where $c$ is a properly chosen constant.
As initial guess in the iteration procedure, we use the static spherically symmetric
$SU(2)$ regular solutions in fixed AdS background \cite{Radu:2001ij}.
The typical relative error for the gauge functions is estimated to be 
lower than $10^{-3}$, while the relations 
(\ref{electric-mass1}) and (\ref{totalJ}) are verified with a very good accuracy. 
For all the solutions we present here
we consider a cosmological constant $\Lambda=-3$.
However, a similar general behavior has been found 
for other negative values of $\Lambda$. 
Similar to the spherically symmetric case \cite{Hosotani:2001iz},
solutions with different values of
$\Lambda$ are related through a scaling transformation.

The solutions depend on two continuos parameters: the values $\omega_0$
of the magnetic potentials $H_2,~H_4$ at infinity and the magnitude of the
electric potential at infinity $V$.
Similar to the static case, the functions $H_2$ and $H_4$ are nodeless or present one node only,
although they have a small $\theta$ dependence. 
For a given $\Lambda$, we found nontrivial rotating solutions for every value of 
$\omega_0$.
A nonvanishing $V$ leads to rotating regular configurations, with nontrivial functions
$H_1,~H_3,~H_5,~H_6$.
As we increase $V$ from zero while keeping 
$\omega_0$ fixed, a branch of solutions forms.
This branch extends up to a maximal value of $V$, which depends slightly on $\omega_0$.
Along this branch, the total energy, electric charge, 
electric part of energy and the absolute value of the angular momentum  
increase continuously with $V$.
Depending on $V$, the energy of a rotating solution can be several order of 
magnitude greater than the energy of
the corresponding monopole solution. 
We find that both $E_{e}/E$ and $J/E^2$  tend to constant values as $V$ is increased.
At the same time, the numerical errors start to increase, we obtain large values for
both $Q_e$ and $E$, and for some $V_{max}$ the numerical
iterations fail to converge. 
In this limit, the total energy and the electric charge diverge, while the magnetic charge
takes a finite value.
An accurate value of $V_{max}$ is rather difficult to obtain, 
especially for large values of $\omega_0$.
Alternatively, we may keep fixed the magnitude of the electric potential at infinity
and vary the parameter $\omega_0$. 
  In Figure 1 we present the properties of typical branches of solutions for 
a fixed value of $\omega_0$ (Figure 1a) and for a fixed $V$ (Figure 1b).

For all configurations, the energy density $\epsilon=-T_t^t$ of the solutions 
has a strong peak along the $\rho$ axis, 
and it decreases monotonically along the symmetry axis, without being possible to distinguish
any individual component. Equal density contours reveal a torus-like shape of the solutions.
Dyon solutions are found in a good portion of $Q_M-Q_e$ plane.
There are also solutions where $Q_M=0$ and $Q_e \neq 0$.
A vanishing $Q_e$ implies a nonrotating, purely magnetic configuration.
However, we find  dyon solutions with vanishing total angular momentum
($J=0$ for some $\omega_0<1$) which are
not static (locally $T_{\varphi}^t \neq 0$) \cite{foot1}.
Resembling the $\Lambda=0$ MAP solutions \cite{Kleihaus:1999sx, Kleihaus:2000hx}, the modulus of the
electric potential $|A_t|$ possesses two zeros at $\pm z_0$, on the $z$ axis.

\noindent{\textbf{Inclusion of gravity.--~}}
We use a metric form inspired by the asymptotically flat ansatz \cite{Kleihaus:2000kg}, 
which satisfies also the circularity condition  \cite{wald}
\begin{eqnarray}
ds^2= \frac{m}{f} ( \frac{d r^2}{1-\frac{\Lambda}{3}r^2}+ r^2 d \theta^2 )
+  \frac{l}{f} r^2 \sin ^2 \theta (d\phi+\frac{\omega}{r}dt)^2 - f(1-\frac{\Lambda}{3}r^2) dt^2,
\end{eqnarray}
with $f$, $l$, $m$ and $\omega$ being functions of $r$ and $\theta$.  
To obtain  asymptotically AdS regular solutions
with finite energy density,
the metric functions have to satisfy the boundary conditions 
$f= m= l =1,~\omega=0$
at infinity, and 
$\partial_r f= \partial_r m= \partial_r l= \omega=0$
at the origin. The boundary conditions on the symmetry axis are 
$\partial_\theta f= \partial_\theta m=\partial_\theta l=\partial_\theta \omega=0, $
and agree with the boundary conditions on the $\theta=\pi/2$ axis.
We remove the dependence on the coupling constants $G$ and $e$ from the differential equations
by changing to dimensionless 

\newpage
\setlength{\unitlength}{1cm}
\begin{picture}(4,6)
\centering
\put(2,0){\epsfig{file=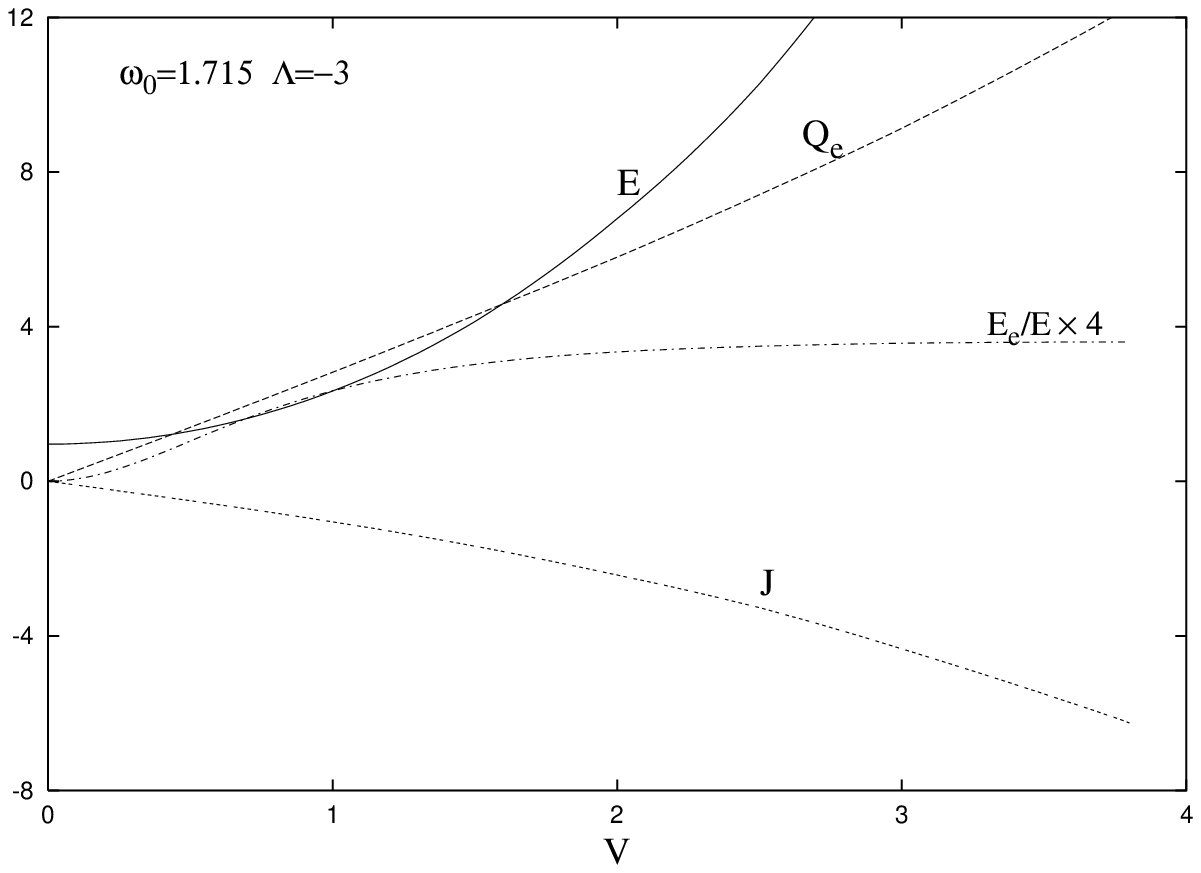,width=10cm}}
\end{picture}
\begin{center}
Figure 1a.
\end{center}
\begin{picture}(20,7.5)
\centering
\put(2.5,0){\epsfig{file=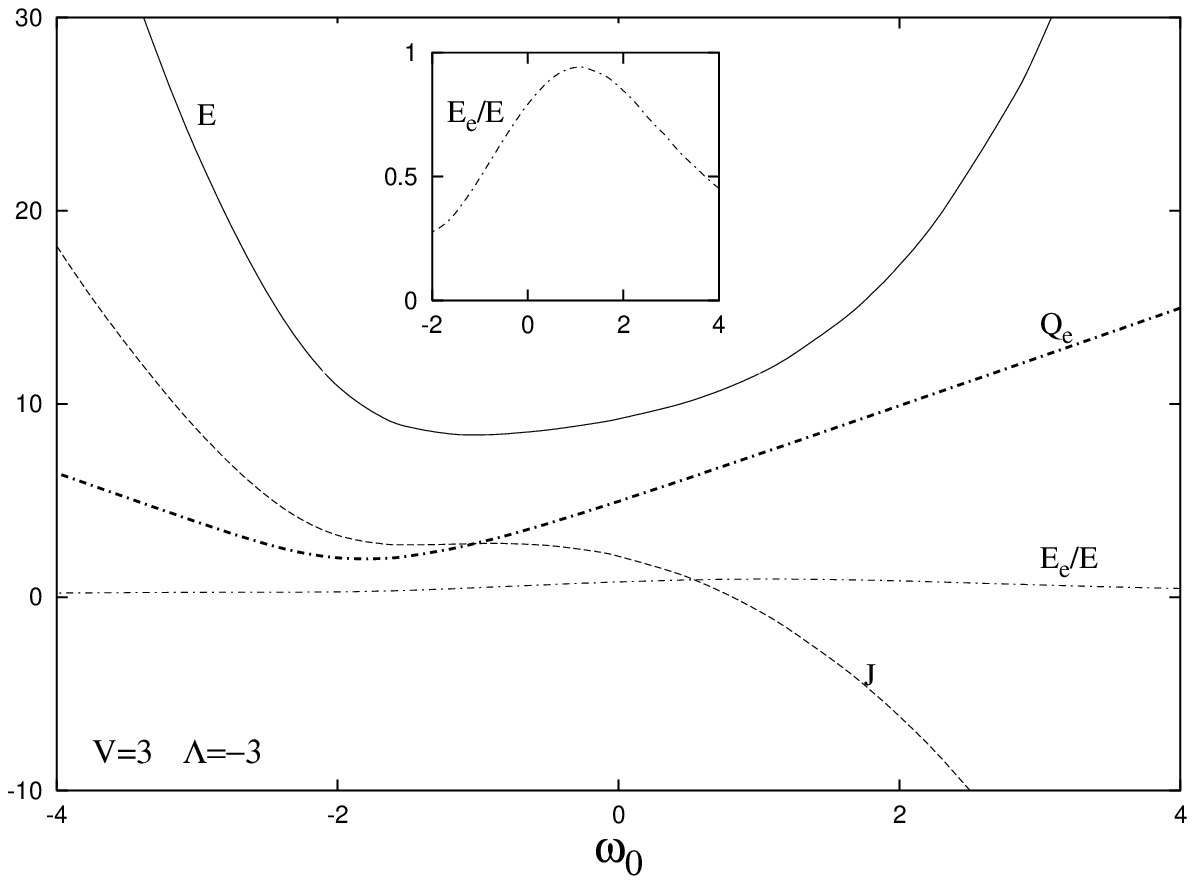,width=10cm}}
\end{picture}
\begin{center}
Figure 1b.
\end{center}
Figure 1. The energy $E$ and the angular momentum $J$ (in units $4\pi/e^2$)
of non-Abelian regular solutions in fixed AdS background with $\Lambda=-3$
are shown as a function on the parameter $V$ (Figure 1a, for $\omega_0=1.715$) 
and the parameter $\omega_0$ (Figure 1b, $V=3$).
Also shown are  the electric charge $Q_e$ and the ratio
$E_{e}/E$.
\\
\\
radial coordinate 
$r \to (\sqrt{4\pi G}/e) r$ and also $\Lambda \to (e^2/4 \pi G) \Lambda$.
The dimensionless mass is obtained by rescaling
$M \to (eG/\sqrt{4\pi G}) M$.
The ADM mass and angular momentum
can be derived from the asymptotics of the metric functions \cite{Abbott:1981ff}.

To solve the EYM equations with a negative cosmological constant
we employ the same numerical algorithm as for the YM solutions in fixed AdS background.
Similar to other  gravitating nonabelian configurations with 
axial symmetry \cite{Kleihaus:1997mn,Radu:2001ij}, 
we use in the numerical procedure a 
suitable combination of the Einstein equations
\begin{equation}
\label{einstein-eqs}
R_{\mu\nu}-\frac{1}{2}g_{\mu\nu}R +\Lambda g_{\mu\nu}  = 8\pi G T_{\mu\nu},
\end{equation}
such that the diferential equations for metric variables $(f,m,l,\omega)$ are diagonal 
in the second derivatives with respect to $r$.

\newpage
\setlength{\unitlength}{1cm}
\begin{picture}(4,8)
\centering
\put(2,0){\epsfig{file=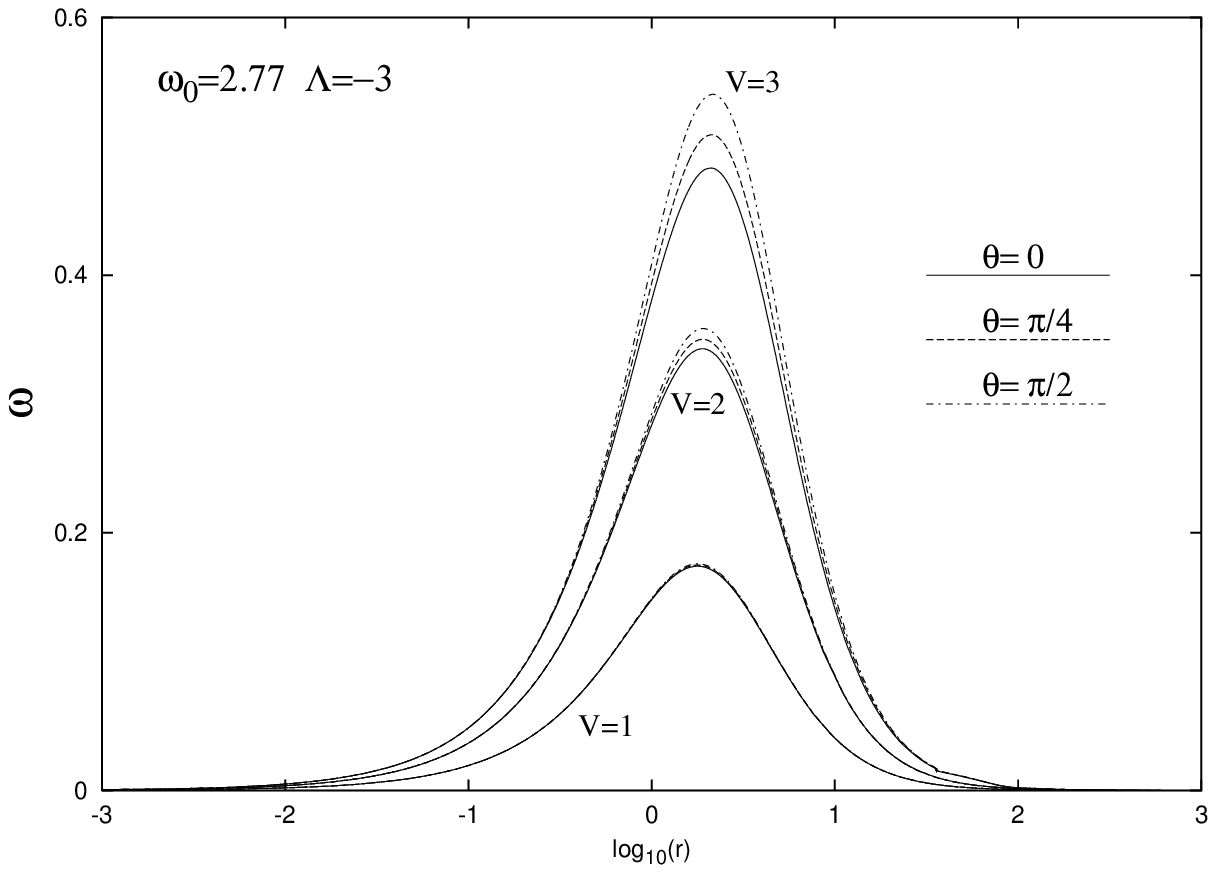,width=12cm}}
\end{picture}
\\
Figure 2. The metric function $\omega(r,\theta)$
is shown as a function of the radial coordinate $r$  
for three fundamental branch solutions
with the same $\omega_0$ and different values of 
the electric potential magnitude at infinity.
\\
\\ 
In the spherically symmetric case, gravitating regular solutions are obtained for 
a finite number of
compact intervals of the parameter $b$.
The properties of these solutions depend essentially on the value of $\Lambda$.
The allowed values of $b$ for YM solutions in an AdS background with a small $|\Lambda|$
correspond approximately 
to the lower branch when coupling to gravity.

We start with a $n=1$ purely magnetic EYM 
solution obtained in isotropic coordinates \cite{Radu:2001ij} as initial guess 
and increase the value of $V$ slowly
(for a fixed $\omega_0$).
We constructed in this way gravitating dyon solutions with nonvanishing angular momentum
for a range of ($\Lambda,~\omega_0,~V$). 
The branch structure of the spherically symmetric solutions is preserved in the presence of rotation.
When gravity is coupled to YM theory, a branch of gravitating dyon solutions emerges
smoothly from the dyon solutions of AdS space and extending up
to some maximal value of $V$ beyond which gravity becomes too strong for regular dyons to persist.
This value is always smaller than the corresponding value in a fixed AdS background.
These lowest branch solutions are of particular interest because some of them are likely 
to be stable against linear perturbations. 
For small values of the cosmological constant (typically $|\Lambda|<10^{-1}$),
the basic properties of these fundamental 
gravitating solutions
are similar to those of the solutions in the absence of gravity.
The general picture presented in Figure 1
provides a good qualitative description in this case too.
We notice that the value at the origin of the metric function $f$ decreases 
with increasing $V$ and tends to zero as 
$V$ approaches the critical value $V_{max}$, corresponding to the formation
of a horizon.

Also, for the studied configurations  we find that the total mass of a gravitating solution 
has a smaller value than the mass of the
corresponding solution in a fixed AdS background.
This inequality is of course in accord with our intuition
that gravity tends to reduce the mass. A similar property has been noticed 
for monopole solutions 
in a spontaneously broken gauge theory \cite{Lee}.
As expected, the gauge functions $H_{i}$ 
look very similar to those of the 
corresponding (pure-) YM solutions.
With increasing $V$, the dyon becomes more and more deformed.
The form of the non-diagonal metric function $\omega$ clearly demonstrates
the differential rotation of different portions of these torus-like configurations (see Figure 2).

Beside these fundamental gravitating solutions, EYM theory with $\Lambda<0$
may possess also excited solutions not presented in a fixed AdS background.
These solutions are obtained by starting with an excited- sherically symmetric 
EYM monopole solution \cite{Bjoraker:2000qd}, and slowly increasing the value of $V$.
We motice the existence of higher node configurations in this case,
the maximal node number depending on $\Lambda$.

For higher branches solutions, we find the same general picture.
The mass, angular momentum and electric charges increase with $V$
and we find again a maximal value for the magnitude of the electric potential at infinity.
Again, as the critical value of $V$ is approached, 
the metric function $f$ develops a zero at the origin,
corresponding to the formation of a horizon.
Also, we find that the metric functions ($f,~l,~m$) of the excited EYM solutions are
considerably smaller at the origin, and the gauge field functions
have their peaks and nodes shifted inwards,
as compared to the corresponding first branch solutions.
Higher branches rotating solutions are more difficult to obtain.
The numerical error for gravitating configurations is estimated to be 
on the order of $10^{-3}$ for first branch solutions and $10^{-2}$ 
in rest.
This error depends also on the values of $\omega_0$ and $V$. 

However, for large values of $|\Lambda|$, the picture presented in 
Figure 1b is no longer valid even for fundamental gravitating solutions. 
For example, we find in this case a minimal allowed value for the parameter $\omega_0$ 
and the absence of higher branches solutions.
These differences  emerge from the distinctions existing already 
in the spherically symmetric case (with $V=0$),  distinctions not yet discussed  in the literature.

More details on these rotating regular solutions
will be given elsewhere.

\noindent{\textbf{Further remarks.--~}}
We found that the gravitating rotating solitons depend nontrivially on the value 
of the cosmological constant $\Lambda$.
As discussed in \cite{Bjoraker:2000qd,Hosotani:2001iz}, 
the higher branches of gravitating monopoles with $\Lambda<0$
are related to Bartnik-McKinnon (BM) solutions \cite{Bartnik:am}.
Naively, one may expect that the corresponding rotating
dyon configurations are also related
to rotating generalizations of BM solutions. In this limit the asymptotics of the electric potential
simplifies: $c_1=c_3=0,~c_2=c_4$ and therefore $J/Q_e=const.$ as predicted in \cite{Brodbeck:1997ek}.
Nevertheless, we find numerically that, in the limit $\Lambda \to 0$, 
the maximal value of $V$ also tends to zero 
and, for higher branches, we recover the nonrotating BM solution
(while the fundamental branch approaches the vacuum solution).
Although further research is clearly necessary, it seems that for 
an axially symmetric regular configuration
with $\Lambda=0$, 
similar to the spherically symmetric case, 
the electric part of the axially symmetric gauge fields is forbidden 
if the ADM mass is to remain finite.
If $V \neq 0$, the $A_t^a$ components of the gauge field act like an isotriplet Higgs field
with negative metric, and by themselves would cause the other components of the gauge field
to oscillate as $r \to \infty$ \cite{VanderBij:2001nm}.

A simple way to remedy this fact is to include a triplet Higgs field in the theory.
The Higgs field functions will satisfy a set of boundary conditions 
similar to the corresponding electric YM potentials.
However, the magnitude of the Higgs field at infinity should be greater that $V$.
This configuration will correspond to an asymptotically flat, rotating MAP 
solution first predicted in \cite{Hartmann:2000ja}, 
with a vanishing net magnetic charge but with a nonzero electric charge.
In this case too, the electric charge and the angular momentum 
are not independent quantities any longer \cite{VanderBij:2001nm}. 

Also, the existence of other branches of asymptotically AdS  rotating regular solutions,
not necessarily connected to the static solutions might be possible.
\\
\\
{\bf Acknowledgement}
\\
The author is grateful to Prof. J.J. van der Bij 
for useful discussions.
The professional observations of the anonymous referee are also acknowledged.
\\
This work was performed in the context of the
Graduiertenkolleg of the Deutsche Forschungsgemeinschaft (DFG):
Nichtlineare Differentialgleichungen: Modellierung,Theorie, Numerik, Visualisierung.


\begin{thebibliography}{99}
\bibitem{foot0}
As $\theta \to \pi-\theta$ we have $H_1\to -H_1,~H_2\to H_2,
~H_3\to -H_3,~H_4\to H_4,~H_5\to H_5,~H_6\to -H_6$
which implies the vanishing of the
integral (\ref{totalJ}). 
The same result is obtained by using the fact that, as $r \to \infty$,  
$H_5\sim V+c/r$ while $H_1$ and $H_6$ vanish faster than $1/r$.
\bibitem{foot1}
From (\ref{totalJ}) we find  
$J=4\pi/3\big(-2c_1/5-c_2+2\omega_0(4c_3/5+c_4)\big)$, 
where $c_i$ are the constants appearing in the 
asymptotics of the electric potential.
\end{thebibliography}
\end{document}